\documentclass[
    aps,
    10pt,
    prab,
    twocolumn,
    amsmath,
    superscriptaddress,
    preprintnumbers,
    floatfix,
    nofootinbib,
    ]{revtex4-2}

\usepackage[utf8]{inputenc}
\usepackage[detect-all,print-unity-mantissa = false]{siunitx}  
\usepackage{filemod}
\usepackage{graphicx}
\usepackage[monochrome]{color}
\usepackage{wrapfig}
\usepackage{amsmath}
\usepackage{bbold}
\usepackage{nicefrac}
\usepackage{soul}
\usepackage[normalem]{ulem}
\usepackage{booktabs}    %
\usepackage{siunitx}     %
\usepackage{caption}     %
\usepackage{multirow}    %
\usepackage{graphicx}    %
\usepackage{subcaption}  %
\usepackage{color}

\allowdisplaybreaks
\usepackage{hyperref}
\usepackage{cleveref}

\begin{document}

\title{Machine learning-based prediction of magnet errors in storage ring light sources}

\author{Jianhao Xu}
\email{xujianhao@ustc.edu.cn}
\affiliation{%
 National Synchrotron Radiation Laboratory, University of Science and Technology of China, Hefei 230029, China
}%

\date{today}

\begin{abstract}
Magnet errors in storage rings significantly degrade beam performance, impacting the brightness and stability of the light source. Therefore, beam-based correction is crucial for the safe operation of machines and the stability of radiated photons. Unlike traditional correction methods such as linear optics from closed orbit, this paper proposes a machine learning (ML) framework to directly predict quadrupole/sextupole gradient errors and misalignment from beam position monitor-measured optics functions and closed-orbit distortion data. Based on a four-bend achromat storage ring lattice, we generate training datasets through ELEGANT numerical simulations and compare regression performance of Linear Regression, Support Vector Machine, Radial Basis Function Neural Network \textcolor{red}{and Densely Connected Convolutional Network}. Results demonstrate that ML models can effectively predict magnet errors and reconstruct ideal optics. This approach offers a novel strategy for accelerating storage ring commissioning and optimization, online diagnostics, and dynamic compensation for next-generation diffraction-limited rings.
\end{abstract}

\maketitle
\section{Introduction}
Beam performance in synchrotron light sources critically determines the  brightness of radiated photons and operational stability. However, magnet errors caused from manufacturing tolerances, misalignment, and power supply noise can distort linear optics and induce closed-orbit distortions (COD), ultimately reducing beam lifetime and source quality\cite{bartolini2008analysis, martin2022orbit, dai2003study, khan2023review, wang2021advances}. Therefore, rapid diagnosis and compensation of magnet errors is a fundamental challenge in accelerator physics.

Traditional correction methods depend on beam response matrix measurements and optimization strategies, such as singular value decomposition (SVD) for orbit correction and linear optics from closed orbit (LOCO) techniques\cite{tsai2006closed, li2001closed, safranek1997experimental, benedetti2011loco}. By measuring the BPM response matrix and iteratively adjusting the magnet current and the strength of the corrector using LOCO, the ideal linear optics can be restored. Since the proposal of LOCO, it has been widely used for optics correction of storage ring light sources. However, this process requires repeated beam experiments, and when there are many parameters, there may be defects such as low accuracy and slow convergence. Crucially, it fails to identify specific magnet errors which caused these perturbations.

Recent advances in machine learning offer new paradigms for orbit and optics correction. Neural networks have been applied in particle accelerator control, beam measurement, commissioning and optimization, BPM detection and prediction, and so on\cite{edelen2016neural, scheinker2017minimization,fol2020detection, veglia2024neural, wang2024bpm,  physrevaccelbeams.28.042801,  scheinker2018demonstration,fol:ipac2022-mopopt047,article}. Compared to conventional methods, ML-based approaches typically enable faster correction with reduced optics distortion. In addition, there have been studies using ML for quadrupole error prediction\cite{fol2021supervised,physrevaccelbeams.26.064601,physrevaccelbeams.28.024601} or time-varying error evaluation\cite{physrevaccelbeams.28.012801}. These studies provide new possibilities for magnet errors prediction and optics correction of storage ring light sources.

Differentiating from ML-based optics correction methods, our framework establishes a direct high-dimensional nonlinear mapping between BPM data and magnet errors, bypassing matrix inversion and error separation. Using a four bend achromat (4BA) storage ring lattice\cite{xu_2025}, we develop an ML-based multi-target regression framework. Training datasets—generated by simulating diverse errors—incorporate linear optics and COD data from BPMs. This enables simultaneous prediction of quadrupole/sextupole gradient errors and misalignments. Our approach provides explicit error identification, reconstructs physical error models, avoids experimental complexity of response matrix measurements, and reveals physical relationships between error types and optics responses—establishing foundations for online diagnostics and dynamic compensation.

This paper is structured as follows: Section 2 details the prediction framework and simulation setup. Section 3 presents data analysis and ML model selection. Section 4 discusses results, and the conclusion is given in Section 5.

\section{METHODOLOGY}
\subsection{Prediction framework}
\begin{figure}
    \centering
    \includegraphics[width=0.7\columnwidth]{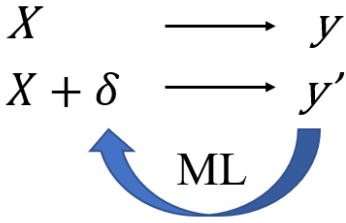}
    \caption{Error prediction framework: from physical modeling to ML-based inverse mapping.}
    \label{figure1}
\end{figure}

The error prediction framework integrates physical modeling and data-driven approaches, as shown in Fig.~\ref{figure1}. Under ideal conditions, particle dynamics follow a deterministic mapping \( X \to y \), where \( X \) represents ideal lattice and magnet parameters and \( y \) represents ideal optics/orbit. Introducing magnet errors \( \delta \) transforms the system to \( X + \delta \to y' \), where the observed data \( y' \) contains distortions caused by \( \delta \). To establish an inverse mapping from observables to error sources, we construct an ML model that learns the nonlinear relationship between parameters \( y' \) and magnet errors \( \delta \). This collaborative framework combines forward physics models with inverse data-driven mapping, enabling rapid magnet error prediction.

\subsection{Data generation and preprocessing}
\begin{figure}
    \centering
    \includegraphics[width=\columnwidth]{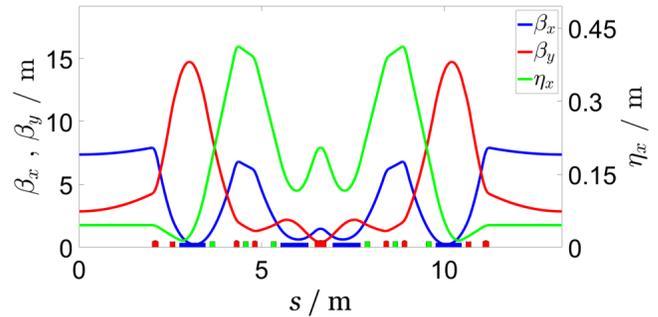}
    \caption{ The magnet layout and linear
optics functions of one cell.}
    \label{figure2}
\end{figure}

\begin{figure}
    \centering
    \includegraphics[width=\columnwidth]{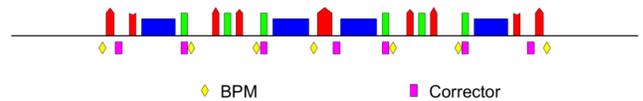}
    \caption{Layout of BPMs and correctors in one cell.}
    \label{figure3}
\end{figure}

A 4BA storage ring designed for HLS-II upgrade\cite{xu_2025} was adopted as the test case. The proposed storage ring lattice has 5 identical cells and the magnet layout and linear optics functions of one cell are shown in Fig.\ref{figure2}, and the BPMs and correctors layout are shown in Fig.\ref{figure3}. Each cell includes 9 quadrupoles (named Q1-Q9 from left to right), 6 sextupoles (named S1-S6), and 7 BPM (named M1-M7). 

The COD of storage rings comes from the dipole field error of various magnets and the misalignment of various magnets. The quadrupole field gradient error and feedback effect caused by the misalignment of strong sextupoles will lead to significant distortion of the optics function. In order to simplify the research, this paper only simulated the gradient error and misalignment of the quadrupoles and sextupoles, with the level of errors referenced in Table.\ref{tab:error_budgets}. All error parameters are generated using a 3\( \delta \) truncated Gaussian distribution to ensure compliance with the actual engineering tolerance range. \textcolor{red}{ELEGANT was performed for simulation and 100000 samples were obtained for training and 10000 samples were obtained for testing\cite{borland2000elegant}.} Here we assume that the corrector was not involved in the simulation, and the optics function and COD at BPM can be directly obtained.

When constructing the dataset, the linear optics (horizontal/vertical $\beta$ function ($\beta_x$, $\beta_y$), dispersion function ($\eta_x$/$\eta_y$)) and closed orbits ($x$, $y$) at all BPMs are used as input features. The input features are named \texttt{Si\_Mj\_$\beta_{x/y}$}, \texttt{$\eta_{x/y}$}, \texttt{$x/y$}, representing the linear optics and orbit at the $j$-th BPM of the $i$-th cell, forming 210-dimensional vectors. Define the error parameters of each magnet as the target variables, which include the gradient error and transverse and longitudinal alignment errors of the quadrupoles and sextupoles. The target variables are $5 \times (9 + 6) \times 4 = 300$ dimensions and are named \texttt{Si\_Qj/Sj\_DX/DY/DZ/FSE}, indicating the different error sources of the $j$-th quadrupole/sextupole of the $i$-th cell. Then Z-score normalization\cite{patro2015normalization} was performed on the feature variables to avoid dimensional differences affecting the accuracy of the model. \textcolor{red}{This process adjusts data values based on how far they deviate from the mean, measured in units of standard deviation.}

\begin{equation}
Z = \frac{X - \mu}{\sigma}
\label{eq:zscore}
\end{equation}
\textbf{Where:}
\begin{itemize}
    \item \(X\) is the value of the data point
    \item \(\mu\) is the mean of the dataset
    \item \(\sigma\) is the standard deviation of the dataset
\end{itemize}

\begin{table}[h]
\caption{RMS error values used for the simulation. The DX/DY/DZ/FSE represents horizontal/vertical/longitudinal misalignment and fractional strength error, respectively.}
\label{tab:error_budgets}
\centering
\begin{tabular}{l c c}
\toprule
& \multicolumn{2}{c}{Magnet Type} \\
\cmidrule(lr){2-3}
Parameter  & Quadrupole & Sextupole \\
\midrule
DX / DY  $(\si{\micro\meter})$  & 100 & 100 \\
DZ $(\si{\micro\meter})$  & 200 & 200 \\
FSE ($ \times 10^{-3} $)  & 1 & 1 \\
\bottomrule
\end{tabular}
\end{table}

\section{Data Analysis}
From the perspective of physical models, magnet errors directly affect the trajectory and optical parameters of the beam. Therefore, when conducting data analysis, the magnet errors are first as features, while the corresponding linear optics functions and COD are used as targets, which is opposite to the prediction model in this paper. By analyzing these high-dimensional data, the correlation between magnet errors, beam optics and COD can be discovered, providing data support for the prediction and compensation of magnet errors. The Pearson correlation coefficient and mutual information coefficient (MIC) \cite{paninski2003estimation} between the magnet errors and the targets were calculated. Pearson correlation coefficient is a statistical indicator that expresses the degree and direction of linear relationship between two variables (X, Y). The calculation formula is:
\begin{equation}
\rho_{X,Y} = \frac{\mathbb{E}[(X - \mu_X)(Y - \mu_Y)]}{\sigma_X \sigma_Y}
\end{equation}
\textbf{Where:}
\begin{itemize}
    \item \(\mathbb{E}\): Expectation operator
    \item \(\mu_X = \mathbb{E}[X]\), \(\mu_Y = \mathbb{E}[Y]\): Means
    \item \(\sigma_X = \sqrt{\mathbb{E}[(X - \mu_X)^2]}\), \(\sigma_Y = \sqrt{\mathbb{E}[(Y - \mu_Y)^2]}\): Standard deviations
\end{itemize}
The MIC is more general and measures the reduction of uncertainty between two variables (X, Y). Compared to correlation analysis, mutual information is non-linear and can capture any form of relationship between variables (linear, non-linear, and even complex dependency structures). The calculation formula is:
\begin{equation}
I(X;Y) = \iint\limits_{\mathcal{X} \times \mathcal{Y}} p_{X,Y}(x,y) \log \frac{p_{X,Y}(x,y)}{p_X(x)  p_Y(y)}  dx  dy
\end{equation}
\textbf{Where:}
\begin{itemize}
    \item \(p_{X,Y}(x,y)\): Joint distribution
    \item \(p_X(x)\), \(p_Y(y)\): Marginal distributions
    \item \(\mathcal{X}, \mathcal{Y}\): Sample spaces
\end{itemize}

\begin{figure*}
    \centering
    \includegraphics[width=\textwidth]{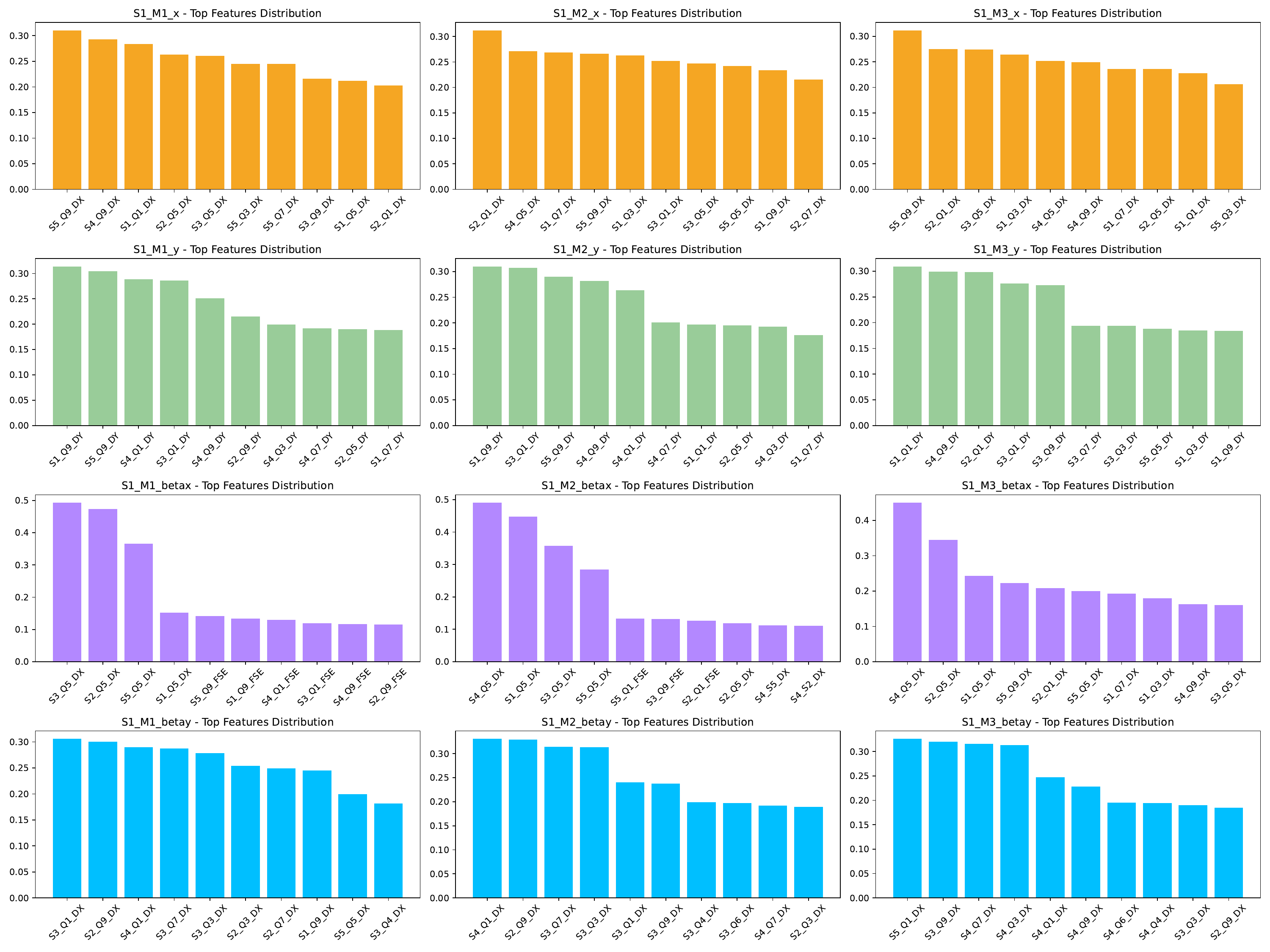}
    \caption{Top 10 most correlated features for representative targets (M1-M3 in Cell 1).}
    \label{figure5}
\end{figure*}

The correlation coefficients between the linear optics and COD of the storage ring and the magnet errors are basically (-0.5 0.5), indicating that there is a complex nonlinear relationship between the magnet errors and responses. Nevertheless, by ranking the importance of features through Pearson and MIC analysis, several significant observations can still be drawn. Fig.\ref{figure5} shows the top 10 important feature rankings corresponding to the orbits and optics functions ($\beta_x$, $\beta_y$) of some BPM locations (\texttt{S1\_M1}, \texttt{S1\_M2} and \texttt{S1\_M3}). It can be seen that the DX and DY of the quadrupole mainly affect the horizontal and vertical orbit distortions, respectively. Among them, the DX and DY of the matched quadrupole Q1 and Q9 have the greatest impact on the horizontal and vertical orbit distortions at the BPM of the whole ring. The DX of Q5 has a significant impact on the horizontal $\beta$-beating of the whole ring. In addition, the DY of the matched quadrupoles Q1 and Q9 basically determines the variation of $\eta_{y}$.

\begin{figure*}
    \centering
    \includegraphics[width=\textwidth]{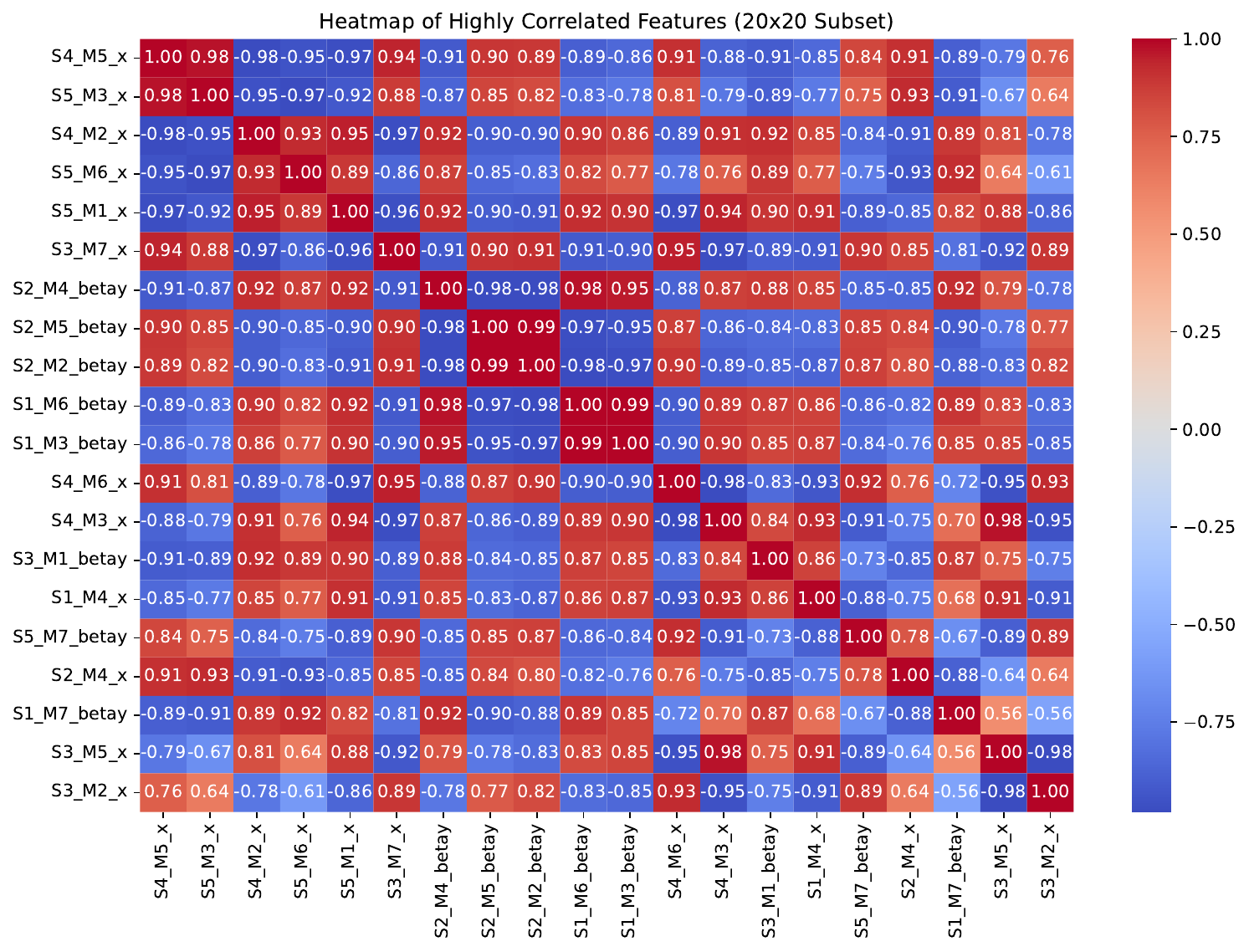}
    \caption{Correlation matrix of BPM-measured parameters.}
    \label{figure6}
\end{figure*}

In addition, we also conducted correlation analysis on 210-dimensional BPM-measured optics functions and orbit. Due to its high dimensionality, 20 targets combinations with the highest sum of absolute correlation values were selected through brute force search algorithm and displayed in the Fig.\ref{figure6}. It can be seen that there is usually a strong correlation between the same physical quantities at different BPMs, indicating that there may be some redundancy between the data. By using dimensionality reduction on data, the data structure can be simplified, and the training efficiency and performance of the model can be improved.

\section{Model Selection and Hyperparameters optimization}
\subsection{Model selection}
In the model selection, we chose four ML supervised models: Linear Regression (LR), Support Vector Machine (SVM), Radial Basis Function Neural Network (RBFNN) and Densely Connected Convolutional Network (DenseNet)\cite{mahesh2020machine,lee1999robust,huang2017densely}. LR serves as the baseline model. SVM regression is a supervised learning algorithm for predicting continuous values. It works by mapping input data into a high-dimensional feature space using kernel functions. The core objective is to find a hyperplane that maximizes the margin while allowing a specified tolerance ($\epsilon$) for prediction errors. RBFNN is suitable for nonlinear problems and function approximation, and its powerful nonlinear mapping ability is expected to play a key role in this study. Below is a brief introduction to RBFNN.

\begin{figure}
    \centering
    \includegraphics[width=\columnwidth]{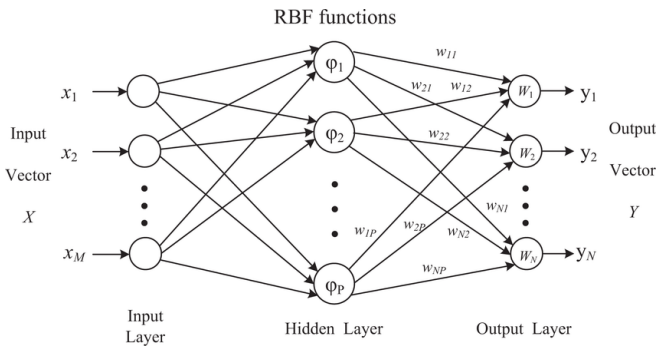}
    \caption{Schematic diagram of RBFNN. \textcolor{red}{A total of 8 models of the DX/DY/DZ/FSE of quadrupoles and sextupoles were trained. For the error targets of quadrupoles, M=210, N=45; for the error targets of sextupoles, M=210, N=30.}}
    \label{fig:RBFNN}
\end{figure}

Radial Basis Function Neural Networks (RBFNNs) represent a class of feedforward neural networks renowned for their efficacy in high-dimensional nonlinear regression tasks. Unlike global approximators (e.g., multi-layer perceptrons), RBFNNs employ localized activation functions centered at prototype points in the input space. This architecture is particularly advantageous for modeling complex physical systems where input-output relationships exhibit strong spatial dependencies, as shown in the Fig.\ref{fig:RBFNN}.
A standard RBFNN comprises three layers:

\begin{enumerate}
    \item \textbf{Input layer}: Receives feature vector $\mathbf{x} \in \mathbb{R}^M$
    
    \item \textbf{Hidden layer}: Computes nonlinear transforms via radial basis functions
    \begin{equation}
    \varphi_i(\mathbf{x}) = \exp\left(-\frac{\|\mathbf{x} - \mathbf{c}_i\|^2}{2\sigma_i^2}\right)\label{equ:Gaussian}
    \end{equation}
    
    where $\mathbf{c}_i$ denotes cluster centers and $\sigma_i$ controls kernel width.
    
    \item \textbf{Output layer}: Generates predictions through linear combination:
    \begin{equation}
    y_k = \sum_{i=1}^{N} w_{ki}\varphi_i(\mathbf{x}) + b_k
    \end{equation}
    
    with weights $w_{ki}$ and biases $b_k$.
\end{enumerate}

\begin{figure}
    \centering
    \includegraphics[width=\columnwidth]{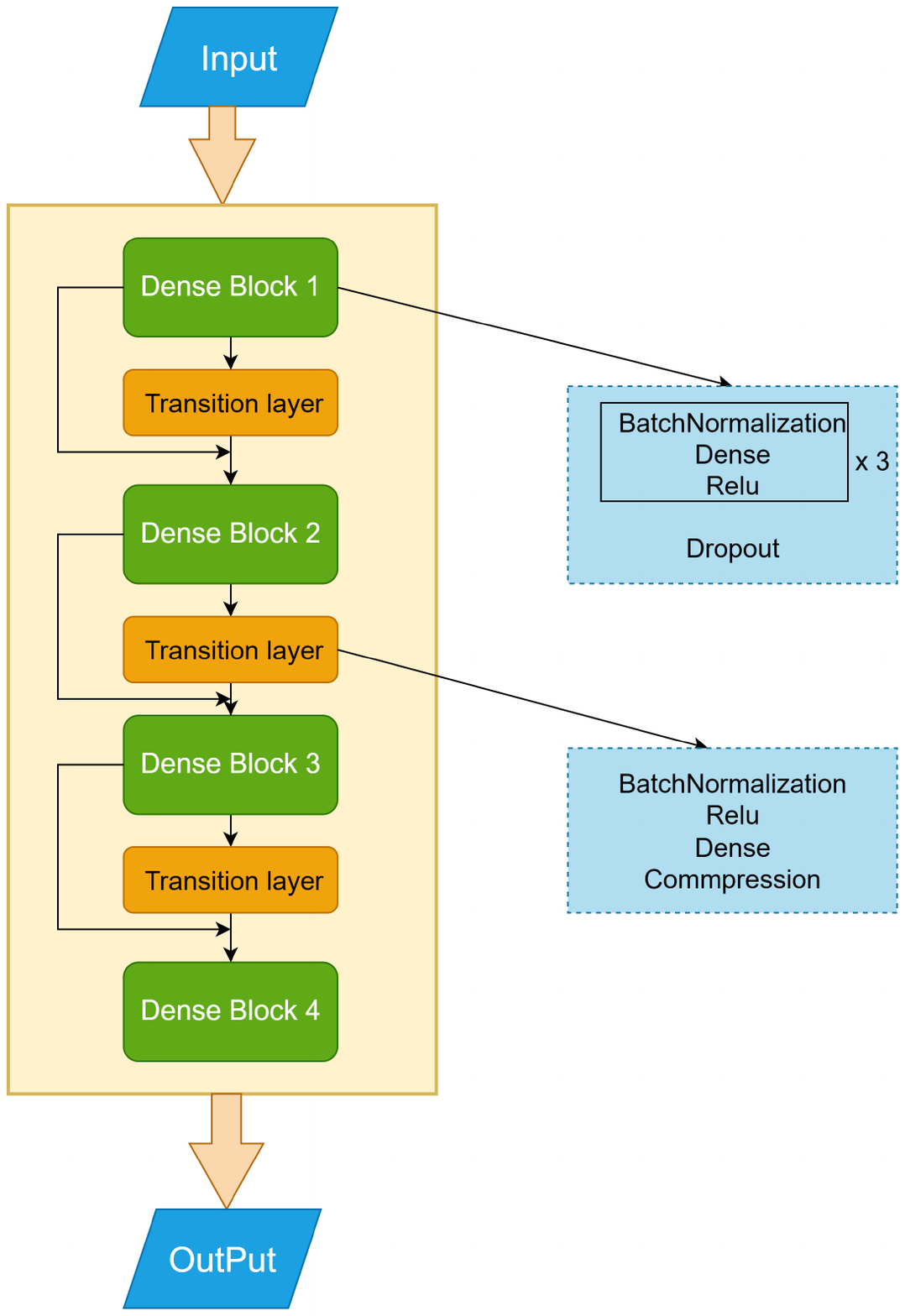}
    \caption{The DenseNet regression model with four dense blocks. The layers between two adjacent blocks are referred to as transition layers.}
    \label{fig:DenseNet}
\end{figure}

The hidden neurons of RBFNN use radial basis functions (such as Gaussian functions, see Eq.(\ref{equ:Gaussian})), and their output depends only on the distance between the input vector and the center point. This local response characteristic allows RBFNN to train without adjusting all connection weights, only through two-stage optimization: first, using unsupervised clustering (such as K-means) to determine the hidden layer center point, and then solving the output layer weights through linear regression, significantly improving the convergence speed and avoiding the problem of local minima.

\textcolor{red}{DenseNet is a prominent deep learning architecture introduced by Huang et al.\cite{huang2017densely}, specifically designed to enhance feature propagation and reuse while mitigating the vanishing gradient problem in very deep networks. Its core innovation is the dense connectivity pattern: each layer receives the feature maps of all preceding layers as input and passes its own feature maps to all subsequent layers within a dense block. This is achieved through direct concatenation, differing significantly from ResNet's summation of residuals.}

\textcolor{red}{Based on the structure of the regression task data and the characteristics of DenseNet, the DenseNet model was modified by replacing the convolutional and pooling layers of dense blocks with fully connected layers. Fig.\ref{fig:DenseNet} shows the architectural details of the DenseNet regression model used in this paper. Four dense blocks were used, with feature compression achieved through transition layers between them, and each dense block was connected through concatenation shortcuts. The output layer is located at the end.}

\subsection{Hyperparameters optimization}
In the stage of hyperparameters optimization, the grid search method was used to optimize the hyperparameters configuration of SVM and RBFNN. For the SVM model, the regularization parameter C was optimized within the range of \( 10^{-2} \) to \( 10^{2} \) to balance the complexity of the model and the fit of the training data. The value of the kernel function $\gamma$ was set between \( 10^{-3} \) to \( 10^{1} \), which affects the range of influence of a single training sample and is crucial for the nonlinear fitting ability of the model. In addition, different settings of $\epsilon$ was explored, ranging from \( 10^{-3} \) to \( 10^{1} \), which defines the regression error tolerance of SVM. The types of kernel functions include linear kernel and polynomial kernel, while the degree of polynomial order considers two possibilities: 2 and 3.

For the RBFNN model, the number of hidden layer center points \( N_{\text{centers}} \) were adjusted between 500 and 2500 to optimize the model's nonlinear mapping ability to input data. The width parameter \(\sigma\) of the radial basis function was set within the range of \( 10^{1} \) to \( 10^{4} \), which controls the local response range of the function and has a significant impact on the sensitivity and generalization ability of the model. Besides, I also tried different types of radial basis functions, including Gaussian function, generalized multi-quadric functions, generalized inverse multi-quadric function and thin plate spline basis function\cite{harpham20062161}, to find the most suitable nonlinear transformation form for this study. \textcolor{red}{The optimal radial basis function for this study is a Gaussian function with \( N_{\text{centers}} \)=2000.}

\textcolor{red}{For this compact 4-layer DenseNet regression model, the hyperparameter optimization was crucial for maximizing the performance. Given the relatively shallow architecture of the model, careful tuning was essential to balance learning ability and prevent overfitting. The key hyperparameters of this model include:
\begin{itemize}
    \item \textbf{Growth rate (\(k\)):} Feature map additions per layer
    \item \textbf{Dropout rate:} Regularization after convolutional layers
    \item \textbf{Batch size:} Gradient stability control
    \item \textbf{Compression rate:} Compressing feature dimensions in the transition layer to control model complexity
\end{itemize}
By randomly generating hyperparameter combinations, the optimal parameters were ultimately determined:
\begin{itemize}
    \item Growth rate (\(k\)) = 16
    \item Dropout rate = 0.3
    \item Batch size = 256
    \item Compression rate = 0.5
\end{itemize}
This combination effectively balanced model complexity and regularization within the 4-layer constraint, with the elevated dropout rate proving particularly valuable for overfitting mitigation.}

The root mean square error ($RMSE$) and coefficient of determination ($R^2$) were used as the performance metrics to evaluate several supervised learning models. In the following formulas, $y_i$ is the true $i$-th value, $\hat{y}_i$ is the predicted $i$-th value, and $\bar{y}$ is the mean of true values.
\begin{equation}
RMSE = \sqrt{\frac{1}{n} \sum_{i=1}^{n} (y_i - \hat{y}_i)^2}
\end{equation}

\begin{equation}
R^2 = 1 - \frac{\sum\limits_{i=1}^{n} (y_i - \hat{y}_i)^2}{\sum\limits_{i=1}^{n} (y_i - \bar{y})^2}
\end{equation}
The smaller the $RMSE$, the closer the $R^2$ is to 1, indicating better predictive performance of the model. By systematically traversing all possible combinations of these hyperparameters and combining cross validation techniques to evaluate the performance of each set of parameters, the hyperparameters configuration with the largest $R^2$ was ultimately determined, ensuring the efficiency and accuracy of the model.

\section{Results and Discussion}
Tables \ref{tab:dx_results}, \ref{tab:dy_results}, \ref{tab:dz_results} and \ref{tab:fse_results} show the results of different ML models on the training and test sets. For different types of errors, DY prediction has the best prediction accuracy. And for different ML models, RBFNN performs the best. When the linear relationship is strong, such as predicting DX and DY of quadrupoles, SVM and RBFNN are almost the same as LR model. However, for strongly nonlinear cases, such as predicting DY and DZ of sextupoles, RBFNN is superior to LR and SVM. All four models exhibited poor performance in predicting the DZ of quadrupoles and the DZ and FSE of sextupoles. \textcolor{red}{DenseNet, while underperforming RBFNN due to preliminary hyperparameters, demonstrates feature-reuse capabilities valuable for future large-scale applications. Its performance is expected to improve with optimized depth and growth rate.}
For the DZ of quadrupole and sextupoles, the impact on linear optics and closed orbit appears significantly weaker than DX/DY and FSE, especially under the assumption of an ideal plane. The poor prediction of the FSE of sextupoles (test $R^2$ = -0.005, see Table \ref{tab:fse_results}) warrants particular attention. This is mainly due to two physical constraints. Firstly, the FSE of sextupole mainly affects chromaticity and nonlinear dynamics, while our input features only include linear optics and COD. Then, the weak correlation between the FSE of sextupole and linear observables suggests that standard regression models cannot extract meaningful patterns.
\begin{table}[h]
\caption{DX prediction results ($R^2$)\label{tab:dx_results}}
\centering
\begin{tabular}{l *{4}{S[table-format=1.4]}}
\toprule
\multirow{2}{*}{Model} & \multicolumn{2}{c}{Quadrupoles} & \multicolumn{2}{c}{Sextupoles} \\
\cmidrule(lr){2-3} \cmidrule(l){4-5}
 & {Training} & {Test} & {Training} & {Test} \\
\midrule
LR & 0.770 & 0.769 & 0.683 & 0.681 \\
SVM & 0.753 & 0.752 & 0.657 & 0.657 \\
RBFNN & 0.779 & 0.777 & 0.722 & 0.720 \\
DenseNet & 0.765 & 0.760 & 0.725 & 0.694 \\
\bottomrule
\end{tabular}
\end{table}

\begin{table}[h]
\caption{DY prediction results ($R^2$)\label{tab:dy_results}}
\centering
\begin{tabular}{l *{4}{S[table-format=1.4]}}
\toprule
\multirow{2}{*}{Model} & \multicolumn{2}{c}{Quadrupoles} & \multicolumn{2}{c}{Sextupoles} \\
\cmidrule(lr){2-3} \cmidrule(l){4-5}
 & {Training} & {Test} & {Training} & {Test} \\
\midrule
LR & 0.857 & 0.857 & 0.753 & 0.753 \\
SVM & 0.848 & 0.848 & 0.750 & 0.750 \\
RBFNN & 0.884 & 0.883 & 0.865 & 0.864 \\
DenseNet & 0.752& 0.740 & 0.680 & 0.651\\
\bottomrule
\end{tabular}
\end{table}

\begin{table}[h]
\caption{DZ prediction results ($R^2$)\label{tab:dz_results}}
\centering
\begin{tabular}{l *{4}{S[table-format=1.4]}}
\toprule
\multirow{2}{*}{Model} & \multicolumn{2}{c}{Quadrupoles} & \multicolumn{2}{c}{Sextupoles} \\
\cmidrule(lr){2-3} \cmidrule(l){4-5}
 & {Training} & {Test} & {Training} & {Test} \\
\midrule
LR & 0.170 & 0.168 & 0.002 & -0.002 \\
SVM & 0.138 & 0.136 & -0.002 & -0.001 \\
RBFNN & 0.225 & 0.221 & -0.004 & -0.004 \\
DenseNet & 0.150 & 0.108 &0.006 &-0.001 \\
\bottomrule
\end{tabular}
\end{table}

\begin{table}[h]
\caption{FSE prediction results ($R^2$)\label{tab:fse_results}}
\centering
\begin{tabular}{l *{4}{S[table-format=1.4]}}
\toprule
\multirow{2}{*}{Model} & \multicolumn{2}{c}{Quadrupoles} & \multicolumn{2}{c}{Sextupoles} \\
\cmidrule(lr){2-3} \cmidrule(l){4-5}
 & {Training} & {Test} & {Training} & {Test} \\
\midrule
LR & 0.617 & 0.615 & 0.002 & -0.002 \\
SVM & 0.591 & 0.588 & 0.016 & -0.035 \\
RBFNN & 0.647 & 0.643 & 0.004 & -0.005 \\
DenseNet & 0.561& 0.510 & 0.003 & -0.001\\
\bottomrule
\end{tabular}
\end{table}

\begin{figure*}[h]
\centering
\begin{subfigure}[b]{0.32\textwidth}
  \centering
  \includegraphics[width=\textwidth]{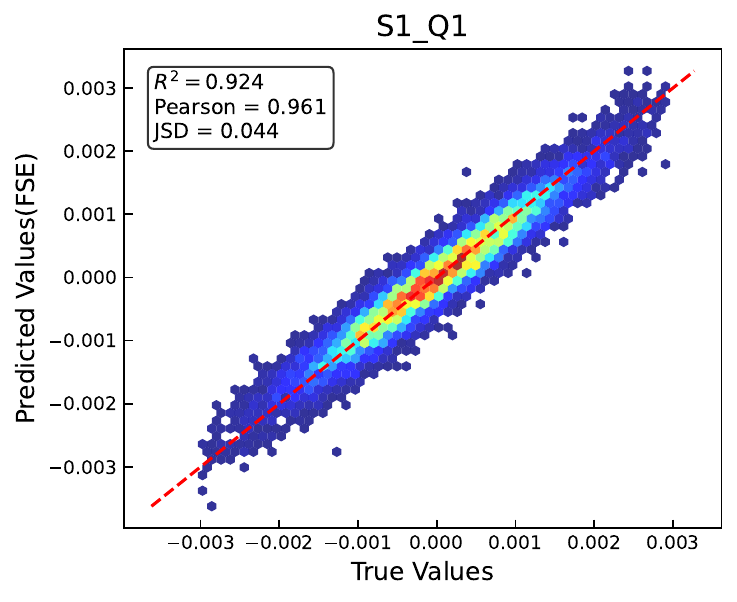}
  \label{fig:fse_s1q1}
\end{subfigure}
\hfill
\begin{subfigure}[b]{0.32\textwidth}
  \centering
  \includegraphics[width=\textwidth]{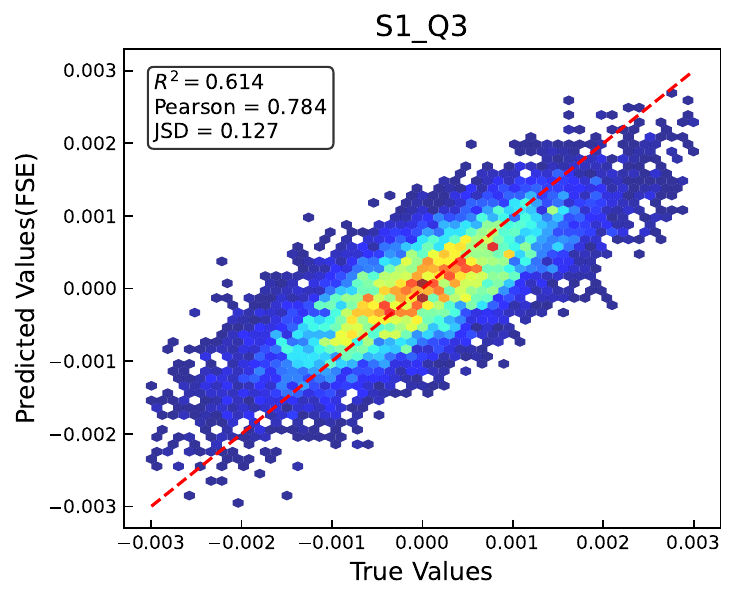}
  \label{fig:fse_s1q3}
\end{subfigure}
\hfill
\begin{subfigure}[b]{0.32\textwidth}
  \centering
  \includegraphics[width=\textwidth]{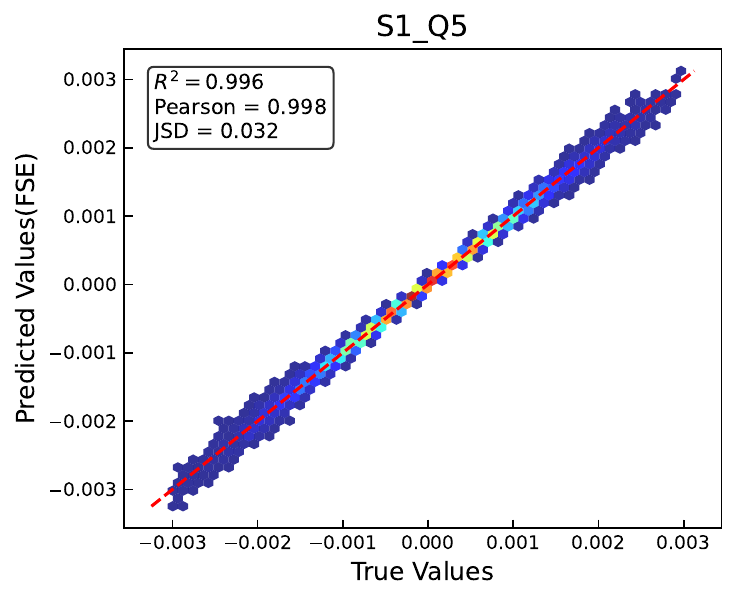}
  \label{fig:fse_s1q3}
\end{subfigure}

\begin{subfigure}[b]{0.32\textwidth}
  \centering
  \includegraphics[width=\textwidth]{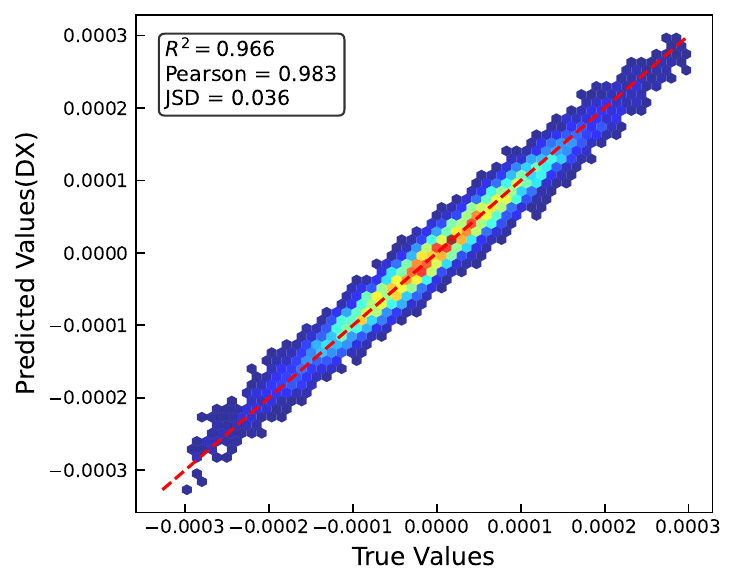}
  \label{fig:dx_s1q1}
\end{subfigure}
\hfill
\begin{subfigure}[b]{0.32\textwidth}
  \centering
  \includegraphics[width=\textwidth]{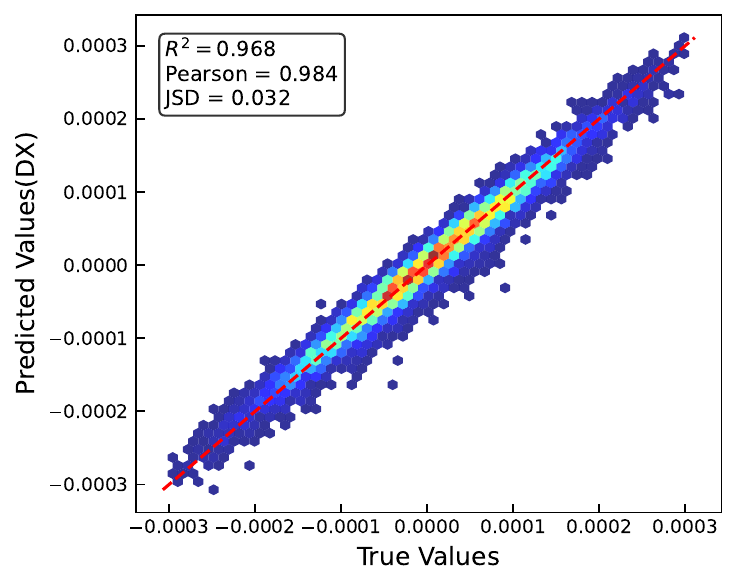}
  \label{fig:dx_s1q3}
\end{subfigure}
\hfill
\begin{subfigure}[b]{0.32\textwidth}
  \centering
  \includegraphics[width=\textwidth]{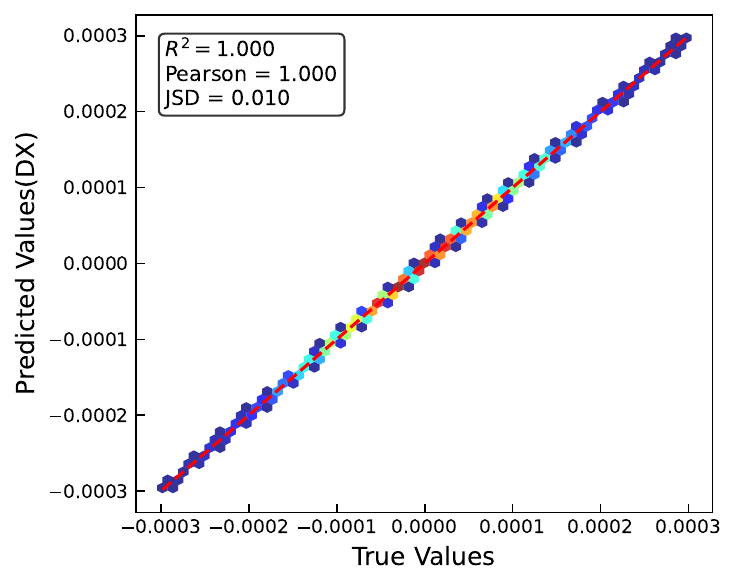}
  \label{fig:dx_s1q3}
\end{subfigure}

\begin{subfigure}[b]{0.32\textwidth}
  \centering
  \includegraphics[width=\textwidth]{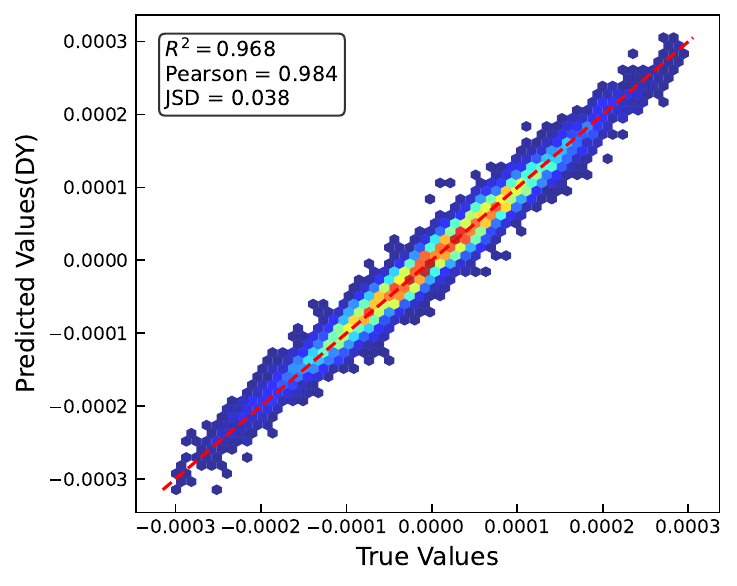}
  \label{fig:dy_s1q1}
\end{subfigure}
\hfill
\begin{subfigure}[b]{0.32\textwidth}
  \centering
  \includegraphics[width=\textwidth]{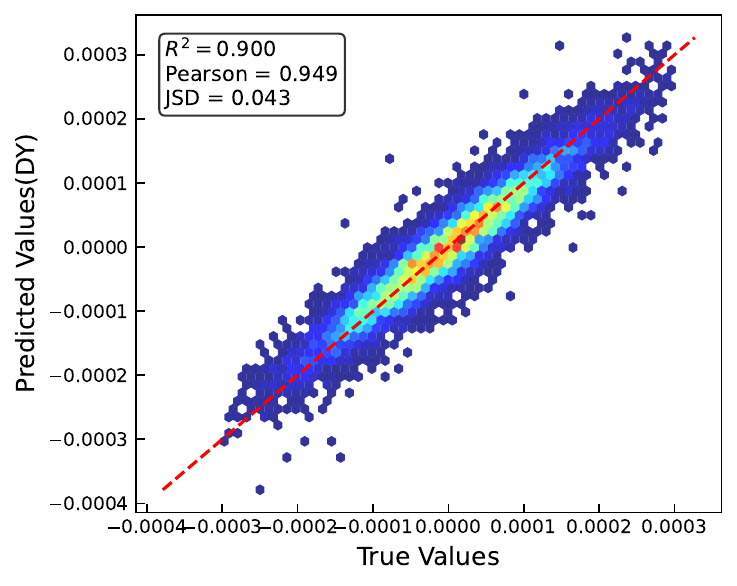}
  \label{fig:dy_s1q3}
\end{subfigure}
\hfill
\begin{subfigure}[b]{0.32\textwidth}
  \centering
  \includegraphics[width=\textwidth]{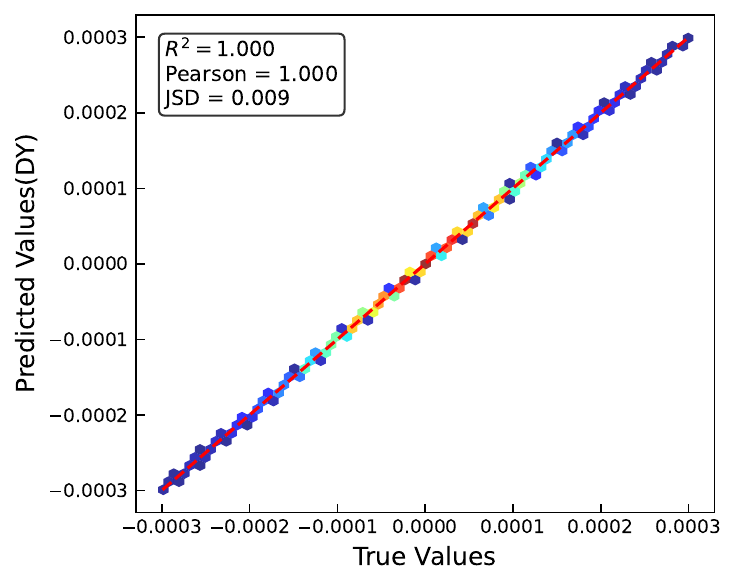}
  \label{fig:dy_s1q3}
\end{subfigure}
\caption{Comparison of true and predicted values of three quadrupoles (\texttt{S1\_Q1}, \texttt{S1\_Q3}, \texttt{S1\_Q5}) on the test set for the LR model. The first to third rows represent three types of errors: FSE, DX, and DY. The colors of the points indicate the density distribution of different data points, ranging from blue (low density) to red (high density), and also represent the Gaussian distribution of the error seeds. The red curve represents the perfect fit line.}
\label{fig:quad_comparison}
\end{figure*}

To further evaluate the model fitting comprehensively, three indicators were used to compare the performance of the prediction model from multiple perspectives: $R^2$, Pearson correlation coefficient, and Jensen-Shannon divergence\cite{menendez1997jensen}. These indicators can effectively evaluate the accuracy and robustness of the model, as well as the similarity between true errors and prediction errors. Fig.\ref{fig:quad_comparison} shows a comparison between the predicted and true errors of three quadrupoles (\texttt{S1\_Q1}, \texttt{S1\_Q3}, \texttt{S1\_Q5}). Among these three quadrupoles, \texttt{S1\_Q5} has the best prediction performance with three target values $R^2 \sim 1$, JSD $<$ 0.05; while the prediction performance of \texttt{S1\_Q3} is poor. This is because Q1 and Q5 are adjacent to BPM, and the features at BPM can better reflect the changes in the two magnets, which also demonstrates the effectiveness of the local correction method. For Q3 (including Q2 and Q4), the prediction performance is worse than Q1 and Q5, which may be due to the stronger disturbance caused by particles after a certain trajectory, resulting in a stronger nonlinear effect between the magnet errors and optics and orbit featuress at BPM.

\begin{figure*}[h]
\centering
\begin{subfigure}[b]{0.32\textwidth}
  \centering
  \includegraphics[width=\textwidth]{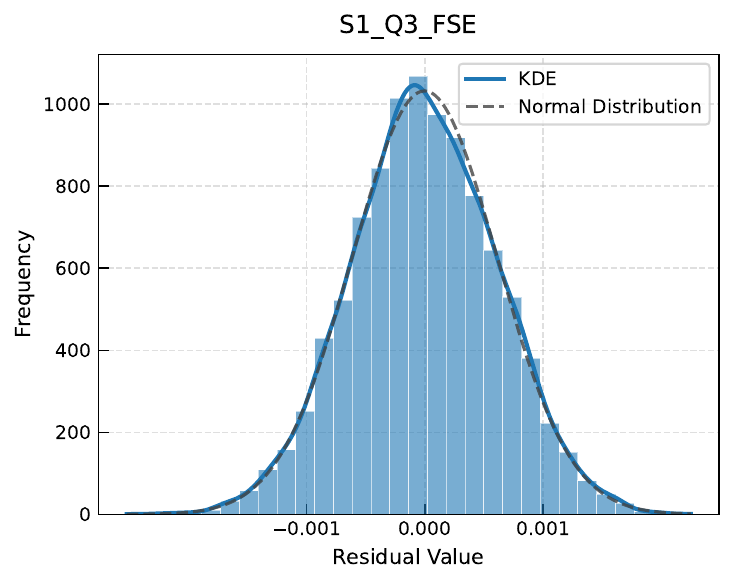}
\end{subfigure}
\hfill
\begin{subfigure}[b]{0.32\textwidth}
  \centering
  \includegraphics[width=\textwidth]{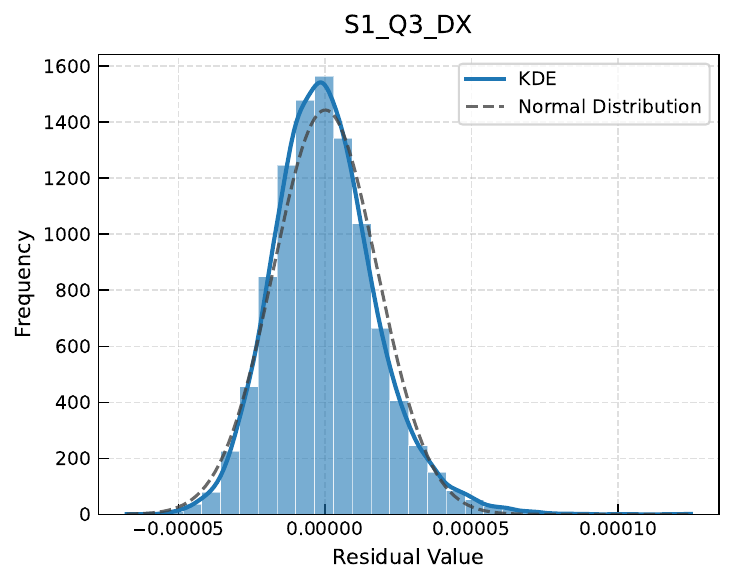}
\end{subfigure}
\hfill
\begin{subfigure}[b]{0.32\textwidth}
  \centering
  \includegraphics[width=\textwidth]{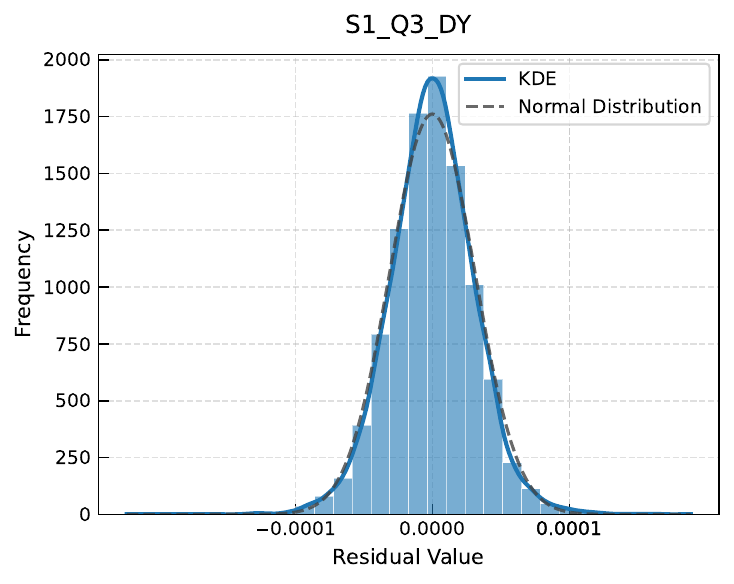}
\end{subfigure}
\caption{The residual distribution of the three types of errors for \texttt{S1\_Q3}, where the residual value is the difference between the true error and the predicted error.}
\label{fig:s1q3_residuals}
\end{figure*}

Then, a histogram of the residual distribution was drawn for \texttt{S1\_Q3}, shown in the Fig.\ref{fig:s1q3_residuals}. The residual histograms of the three errors values all exhibit approximately symmetrical unimodal distributions. The KDE curve (blue) highly overlaps with the theoretical normal distribution (dashed line) in the central region, indicating that the residuals follow the normality assumption near the mean. However, for the DX error, the tail of the residual on the right side is relatively long, and there may be a small number of high-value outliers.

\begin{figure}
    \centering
    \includegraphics[width=\columnwidth]{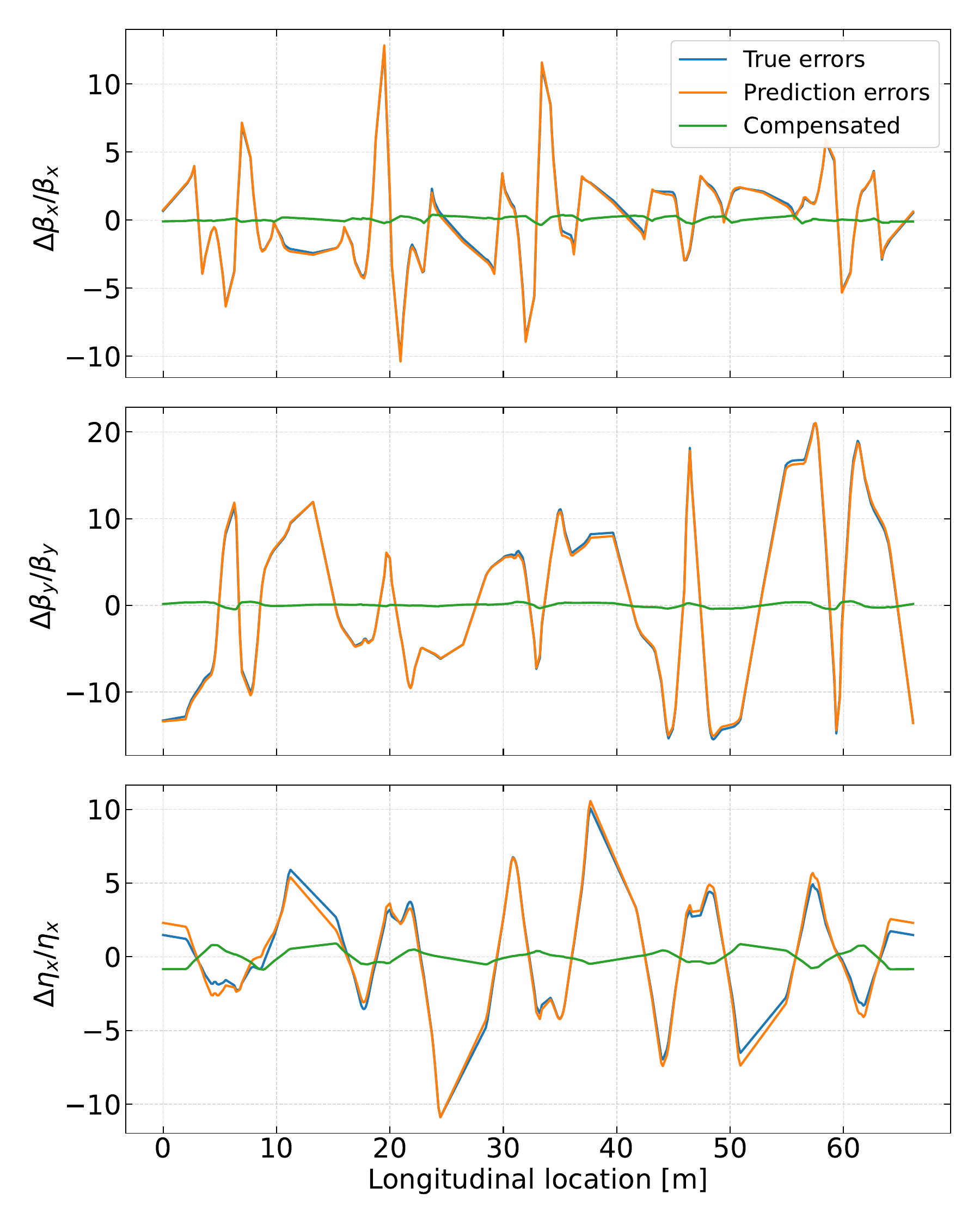}
    \caption{Reduction of $\beta$-beating and $\eta_x$ achieved by compensating predicted magnet errors.}
    \label{fig:compensated_optics}
\end{figure}

\begin{figure}
    \centering
    \includegraphics[width=\columnwidth]{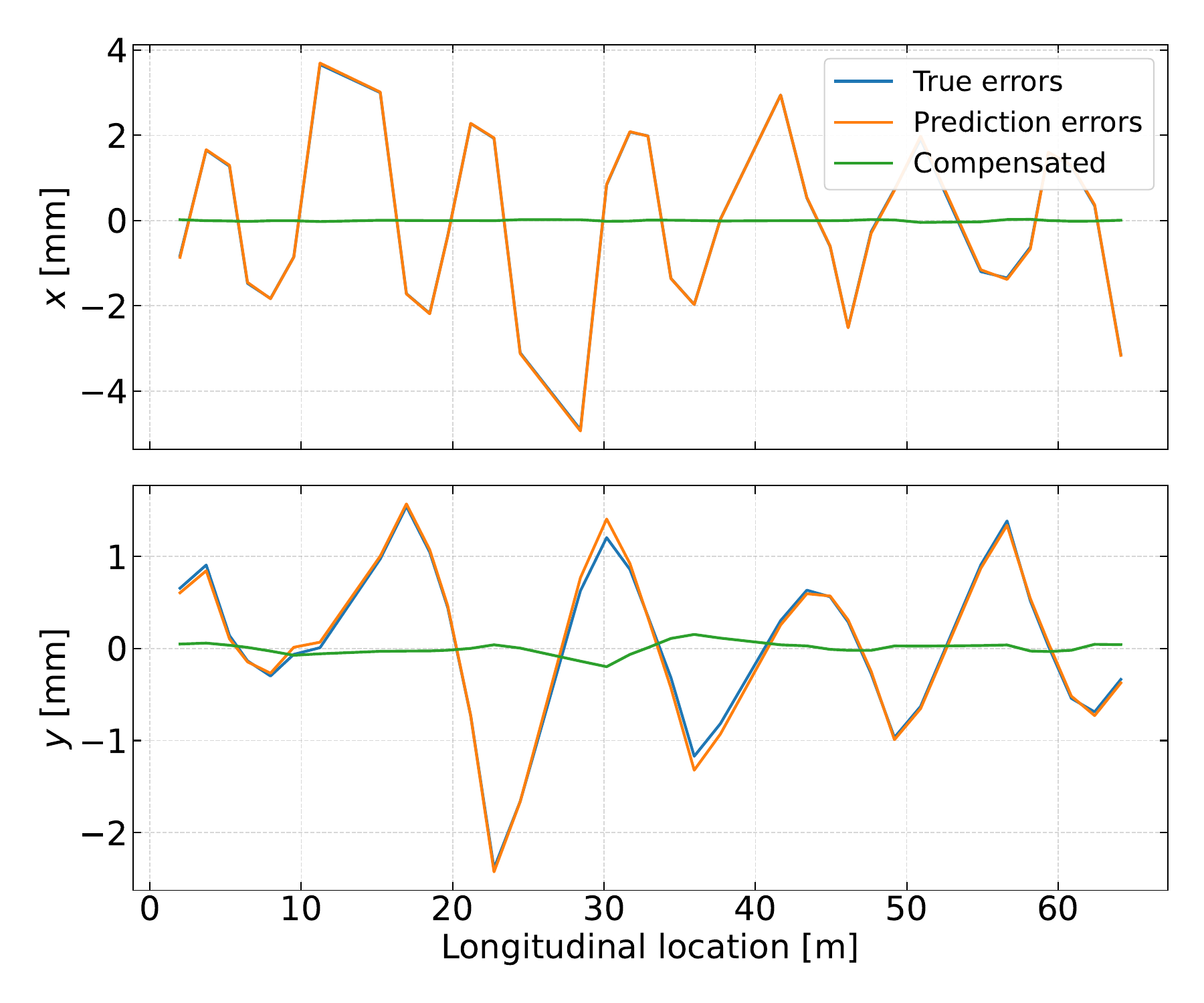}
    \caption{Closed orbits along the ring.}
    \label{fig:orbit}
\end{figure}

In order to test the predictive ability of the trained regression model, a new random error seed was generated to compare the actural simulated magnet error with the error predicted by the ML model. Fig.\ref{fig:compensated_optics} shows the $\beta$-beating of the original simulation error, the $\beta$-beating of the prediction error, and the $\beta$-beating after adjusting the FSE, DX, and DY of the quadrupoles and the DX and DY of the sextupoles to the ideal state (assuming these values can be adjusted). Fig.\ref{fig:orbit} shows the closed orbit along the ring. It can be seen that the difference between the linear optics and orbit distortions caused by the true errors and prediction errors are very small. After compensating for the prediction errors, \textcolor{red}{the RMS horizontal and vertical $\beta$-beating were reduced from 3.22\% to 0.17\%, and from 9.75\% to 0.25\%, respectively. The RMS value of $\eta_x$ varition was reduced from 3.65\% to 0.45\%.} The closed orbits also showed significant improvement after compensation. The adjusted value has reached the level of using conventional correction methods.

\section{Summary and Conclusion}

This paper demonstrates the effectiveness of machine learning in predicting complex magnet errors in a storage ring light source through systematic numerical simulation and multi model comparison. Through the framework proposed in this paper, we can effectively predict magnet errors, which are of great significance for accelerating the commissioning of storage rings, online error diagnosis, and dynamic compensation. For example, by predicting the gradient error and alignment error of a quadrupole and sextupole, we can more accurately adjust the magnet current, thereby optimizing the beam performance. In addition, this paper also explores the correlation between magnet errors, linear optics functions, and COD through correlation analysis and mutual information analysis. These rules not only provide a theoretical basis for the construction of machine learning models, but also provide useful guidance for the physical design and optimization of storage rings. For example, by analyzing the correlation between magnet error and optics functions, it is possible to identify the magnet position and error type that have the greatest impact on beam performance, and thus focus on controlling them during magnet manufacturing and collimation processes. 	

However, there are still some shortcomings in this study. For example, a more comprehensive error setting, including rotation error, high-order field error, etc., were not adopted in the simulation; and the practical application of the correctors has not been considered in this paper. In addition,  for larger-scale fourth-generation storage rings, due to more feature variables and stronger nonlinear effects, applying this framework may require incorporating feature selection or dimensionality reduction techniques to handle high dimensionality, as well as employing more sophisticated deep learning architectures (such as Transformers) to capture long-term dependencies and strong nonlinearities, or developing staged/regional prediction strategies.

\bibliography{literature}

\end{document}